# A typical workflow to simulate cytoskeletal systems with Cytosim.

Carlos A. Lugo*, Eashan Saikia* and Francois Nedelec†
Sainsbury Laboratory, Cambridge University,
Bateman Street, Cambridge CB21LR, United Kingdom.
* Contributed equally, alphabetical order. †Correspondence: fjn28@cam.ac.uk

**SUMMARY:**

This protocol demonstrates how to use Cytosim, an Open Source cytoskeleton simulation, to investigate the behavior of a network of filaments connected by molecular motors and passive crosslinkers. A generic workflow with step-by-step instructions is followed to vary the number of crosslinkers and plot the resulting network contractility.

**ABSTRACT:**

Many cytoskeletal systems are now sufficiently well known to permit their precise quantitative modelling. Microtubule and actin filaments are well characterized, and the associated proteins are often known, as well as their abundance and the interactions between these elements. Thus, computer simulations can be used to investigate the collective behavior of the system precisely, in a way that is complementary to experiments. Cytosim is an Open Source cytoskeleton simulation suite designed to handle large systems of flexible filaments with associated proteins such as molecular motors. It also offers the possibility to simulate passive crosslinkers, diffusible crosslinkers, nucleators, cutters and discrete versions of the motors that only step on unoccupied lattice sites on a filament. Other objects complement the filaments by offering spherical or more complicated geometry that can be used to represent chromosomes, nucleus or vesicles in the cell. Cytosim offers simple command-line tools for running a simulation and displaying its results, that are versatile and do not require programming skills. In this workflow, step-by-step instructions are given to: i) install the necessary environment on a new computer, ii) configure Cytosim to simulate the contraction of a 2D actomyosin network, iii) produce a visual representation of the system. Next, the system is probed by systematically varying a key parameter: the number of crosslinkers. Finally, the visual representation of the system is complemented by a numerical quantification of contractility to view, in a graph, how contractility depends on the composition of the system. Overall, these different steps constitute a typical workflow that can be applied with few modifications, to tackle many other problems in the cytoskeletal field.

**INTRODUCTION:**

The cytoskeleton consists of filaments within the cell and associated molecules such as molecular motors, which are often forming together, a dynamic meshwork with remarkable mechanical properties. The cytoskeleton exists in many flavors in different cell types across nearly all life forms. Its correct functioning is essential for fundamental cellular processes such as division, motility, polarization, etc. It also governs cell-to-cell mechanical interactions, and thereby influences morphogenesis of tissues and organisms. The cytoskeleton underlies myriad of functions and manifests itself in many biological processes. For example, the contraction of muscles is linked to the power stroke of myosin molecular motors on actin filaments. Another example is the maintenance of neurons, which relies on the movements of kinesin motors along microtubules located inside the axons of these neurons. Actin and microtubules are two preeminent type of cytoskeletal filaments, and without them, life as we know it, would be impossible.

The cytoskeleton is by essence a biomechanical system, which cannot be reduced solely to its chemistry. Microtubules or actin filaments are built from thousands of monomers and extend over several micrometers. The conformations of these filaments in space and the forces that they are able to transmit to the plasma membrane, to the nucleus, or to other organelles are key aspects of their role in the cell. For example, a network of actin filaments and myosin motors, called the actomyosin cortex[1], generates forces to sustain cell motility and morphological changes in animal cells. A very different arrangement is seen in plant cells, where cortical microtubules direct the deposition of cellulose fibrils, thereby controlling the cell wall architecture, which ultimately determines how these cells will grow in the future[2]. While mechanics clearly play a substantial part in cytoskeletal operations, chemistry is equally important. Filaments grow via a self-assembly process whereby monomers find their docking site at the tip of the filament after diffusing through the cytoplasm[3]. At the molecular scale, assembly and disassembly at the tip of the filaments are thus determined by molecular affinities[4]. Similarly, proteins of the cytoskeleton diffuse, and binding and unbinding rates determine their affinity for the filaments they encounter. In the case of molecular motors, cycles of chemical reactions involving ATP hydrolysis are linked to movements along the filaments, and possibly forces that accompany them[5]. Remarkably, the cytoskeleton offers many unusual challenges, and a large variety of processes involving similar components. It is a rich playground at the interface between biology, chemistry and physics.

Cytoskeletal systems are amenable to mathematical modelling. In fact, thanks to excellent research done in the past decades, the principal molecular constituents are most likely already identified, as illustrated with endocytosis[6]. In model organisms, such as yeast, the properties of these elements are known, as well as the system composition for some of their processes. For instance, the structure and material properties of microtubules[7], their number and average lengths at various stages of the mitotic spindle have been described[8]. Often known are the number of kinesins that connect microtubules into a coherent mechanical structure[9]. The speeds of many motors have been measured *in vitro*[10]. In addition, experimentalists can observe and quantify these systems *in vivo* under wild-type or mutated conditions. Combining theory alongside *in vivo* and *in vitro* experiments enables researchers to test if the current knowledge about a cytoskeletal system is sufficient to explain its observed behavior. The use of mathematical and computational tools also allows to make inferences of how components work collectively, on the basis of assumptions derived from observations at the molecular scale, usually in simplified situations, *e.g.* single molecule experiments.

Let's illustrate the role of theory using a practical example: the beating of cilia. This beating is due to the movement of dynein motors along microtubules in the cilia. One may ask: what determines the speed of the dynein motor in this system? One possible answer is that the maximum speed is constrained by the requirement to maintain a certain beating pattern. This would be understandable if the beating is under natural selection. In that case, if motors moved quicker, then the process would lose its desired qualities, *i.e.* the cilia would not beat as efficiently, or even fail altogether. Although this is possible, a second alternative is that some intrinsic factor could limit dynein's speed. For example, the cell may not have enough ATP to make dynein faster, or the protein movements required for dynein's activity just could not be accelerated. In that case, if the motors could be made faster despite the physical limits, the beating would be improved. A third possibility, of course, is that changing the speed does not affect the process significantly, which might be advantageous to the organism by providing some 'robustness' against uncontrollable factors. Amongst these three possibilities, one can identify the correct one by calculating the beating pattern from dynein's properties. Indeed, a suitable mathematical model should predict how the beating pattern is affected by varying dynein's speed, and is not subject to the limits that exists in the physical world. Naturally, the validity of the model must be verified, but even 'incorrect' models can generate interesting ideas.

The model can take the form of an analytical framework or be a numerical simulation of the system. Either way, the gap between the molecular scale and the functional scale remains an obstacle, and developing these models is not a straightforward task, since several mechanical and chemical processes need to be integrated into the equations describing the biological system. Theory comes in various flavors, offering different tradeoffs between simplicity and realism. Increasing the degree of details in a model is not always advantageous as it may limit our ability to solve the equations, or in other words, to derive the predictions of the theory. The same tradeoff exists for simulations. The modelers will have to select the aspects of the system to be considered, while ignoring certain aspects. These key decisions will depend strongly on the objective of the study. Nowadays, the extraordinary improvements of computer hardware make it possible to simulate many cytoskeletal systems with enough details over a sufficient time to analyze their comportment. This will often generate unexpected ideas and novel directions in our research. For example simulations similar to the ones that will be used in this protocol lead to a back-of-the-envelope calculation that can predict the contractility of a network based on its composition[11].

Numerical methods are ubiquitous in engineering and physical sciences, and their use in biology is growing. Today virtually all our technological whatchamacallit (watches, phones, cars, computers...) have been first conceived on a computer, and powerful software exist to do this. Given a well-characterized cytoskeletal system and assuming that an appropriate level of description has been determined, several issues must still be solved before it can be simulated. For the simplest problems, the most appropriate route of action might be to write a simulation "by coding from scratch", in other words, starting with a generic programming language or a mathematical platform such as MATLAB. This has the advantage that the author of the code will have an intimate knowledge of what has been implemented and knows exactly how the software works. This route is however not without risk, and it is not uncommon to witness PhD students spending most of their working time writing code rather than addressing scientific questions. The alternative is to use software conceived by others, but this is not without risks either: any large source code tends to spontaneously acquire the traits of an impenetrable black box, despite the most admirable efforts of their authors to prevent it. Using black boxes is surely not a scientist's dream. A large source code can also become a liability, and it may be quicker to start from scratch than to modify an existing code base to make it do something different. To mitigate this problem, one can always invite the authors of the software to help, but this may not be sufficient. Frequently there is a difference of scientific culture between the authors of the software, and the people that would like to use it, which means that many implicit assumptions need to be clarified. By making the code Open Source, it is expected that more people will be involved in the development of the software and to maintain its documentation, thus improving its quality. All these are important issues that must be given proper consideration before any investment is made. Nevertheless, the only way to progress in the long term is to promote solid software solutions, used and maintained by a broad community with common scientific interests.

Although this protocol uses Cytosim, there are other Open Source tools that might be able to simulate the same system, for example AFINES[12], MEDYAN[13], CyLaKS[14], aLENS[15], and AKYT[16], to name a few. Unfortunately, comparing these projects is beyond the scope of the article. Here, step-by-step instructions are given to simulate a contractile 2D actomyosin network. This system is simple and makes use of the better-established capacities of Cytosim. Cytosim is built around a cross-platform core engine that can run simulations in 2D or 3D. It has a modular code base, making it easily customizable to meet particular tasks. Cytosim is equally stable and efficient in 3D, and it has been successfully used in the past to investigate diverse problems involving microtubules and actin filaments: the association of two asters of microtubules[17], the movement of nuclei in the cells[18,19], endocytosis[6], cytokinesis[20], the formation of the mitotic spindle[21], the movements of the mitotic spindle[22], capture of chromosomes[23], contraction of actomyosin networks[11,24], the mechanics of the microtubule ring in blood platelets[25], and the capacities developed for these projects have been maintained in the code. The workflow described here can be adapted to many other problems. It makes use of the Unix command line, which may be unfamiliar to some readers. Using the command line is, however, the most portable and convenient way to automate the process of running simulations. Integrated GUIs aim to offer easy intuitive access to a software, but this often comes at the expense of generality. The objective of this article is to illustrate an approach that can easily be modified or adapted to other problems. Notes are provided as necessary to explain the meaning of the commands.

To simulate an actomyosin network, filaments will be modeled as oriented lines, and represented by vertices distributed along their length (Fig. 1). This is an intermediate level of description, common in polymer physics, that ignores the genuine three-dimensional nature of the filaments, but allows the bending to be calculated. Filaments may grow and shrink at their ends, following different models that cover both actin and microtubule phenomenology. In cells, filaments are organized primarily through interactions that constrain their motion, for example the attachment to other filaments, or simply confinement within the cell. In Cytosim, all such interactions are linearized and combined in a large matrix[26]. The equations describing the motion of all filament vertices are derived from this matrix, assuming a viscous medium and random fluctuating terms representing Brownian motion. These equations are solved numerically to obtain the motion of the filaments together with all the forces acting on them in a self-consistent and efficient manner[26]. Superimposed on this mechanical engine, there is a stochastic engine which simulates discrete events, such as the attachments and detachments of molecular motors or the assembly dynamics of filaments. In summary, Cytosim firstly uses simulated dynamics to calculate the mechanics of a network of filaments, connected in any arbitrary manner and, secondly, stochastic methods to simulate the binding, unbinding and diffusion of proteins that connect or affect the filaments.

The workflow illustrated here was frequently followed to initially explore a system using Cytosim. The critical step for many potential users is likely to be the installation of the software components. Distributing the software as source-code fulfills the imperatives of Open Science, but it is prone to errors since the developers of the software only have access to a limited pool of architecture to test the program. Compilation may fail as operating systems differ. The instructions provided here are likely to become obsolete as computer systems and source codes evolve. Thus, periodically checking the latest instructions online is essential. In case of trouble, it is highly encouraged that users report back by posting on the relevant feedback channel (currently Cytosim's homepage on Gitlab), to give a chance for the problem to be fixed.

The protocol consists of these steps:
    1) Platform preparation, for Windows 10, MacOS and Linux
    2) installation of Cytosim
    3) configuration of the simulation and test run, graphical display
    4) multiple runs, varying a parameter: the number of crosslinkers in the network
    5) generating a graph to view how contractility is affected by the number of crosslinkers

      6) parallel runs
      7) random sampling

Note that all text following a '>' are command lines that are to be entered in the terminal window.
The '>' represents the terminal prompt and must not be included, but all other characters are important.

**Filming will cover parts 3.3 to 3.8, 4.3 to 4.9 and 7 (all parts).**
# 1. Platform preparation
**Depending on the OS (MacOS, Windows 10 or Linux), follow 1.1, 1.2 or 1.3**

## 1.1 - Preparation (MacOS)

**1.1.1 -** Install Xcode Command Line Tools: Open the Terminal (in Applications/Utilities), and enter
```
> xcode-select --install
```

**1.1.2 -** To view or edit Cytosim's code, install Xcode from the Apple App Store.
```
https://apps.apple.com/us/app/Xcode/id497799835?mt=12
```

**1.1.3 -** Xcode's built-in editor is perfectly suitable. Other code-orientated editors would work, for example TextMate. For TextMate, download and follow instructions from https://macromates.com

## 1.2 - Preparation (Windows)

For Windows 10 or higher, Cytosim can run using the "**Windows Subsystem for Linux**" (WSL), as described below. An alternative for older version is Cygwin, but instructions are not provided here.

**1.2.1 - Update** the computer operating system to fulfill requirements for WSL 2:
      Windows 10 version 1903 or higher, with Build 18362 or higher, for x64 systems
      Version 2004 or higher, with Build 19041 or higher, for ARM64 systems.
    For subsequent version updates, please check:
    https://docs.microsoft.com/en-us/windows/release-health/release-information

**1.2.2 -** Type "Turn Windows features on and off" on the search box of the taskbar. Manually enable (☑) Virtual Machine Platform and (☑) Windows Subsystem for Linux. Click OK and Restart Windows.

**1.2.3 -** Go to Windows "Microsoft Store" and search for "Ubuntu". Download and install the current release (Ubuntu 20.04 LTS as of 03.2022)

**1.2.4 -** Click Launch to start the Ubuntu Terminal. If asked to download the latest WSL2 Linux Kernel, proceed by follow the instructions that will be provided. Install the update and relaunch Ubuntu.

**1.2.5 -** Follow terminal instructions to "Enter new UNIX username" and set a password.
Once the user account is set up, Ubuntu can be launched from the search box of the Windows task bar.
Opens the home directory (aka '.'), from the command window:
```
> explorer.exe .
```

**1.2.6 -** Install a WSL-compatible X-Window server, for example *Xming*:

[https://sourceforge.net/projects/xming/](https://sourceforge.net/projects/xming/)

**1.2.7 –** Start the X-Window server, *Xming*:
Double click the Xming icon; select "Multiple Windows"

Hereafter, open the Ubuntu terminal and follow the following steps in **1.3** (Linux)**.**

## 1.3 - Preparation (Linux)

**Note:** these instructions are appropriate for Linux distributions that use the APT package manager. These commands must be modified for distributions such as Red Hat Linux that use a different package manager. In this case, follow the instructions of the Linux distribution to install the same packages.

**1.3.1 -** Update the Linux system
```
> sudo apt-get update
> sudo apt-get upgrade
```

**1.3.2 -** Install a C++ compiler and GNU's make (https://www.gnu.org/software/make)
```
> sudo apt-get install build-essential
```

**1.3.3 -** Install BLAS/LAPACK Libraries (http://www.netlib.org/lapack)
```
> sudo apt-get install libblas-dev liblapack-dev
```

**1.3.4 -** Install OpenGL developer library and header files (https://www.mesa3d.org)
```
> sudo apt-get install mesa-common-dev
```

**1.3.5 -** Install the GLEW library (http://glew.sourceforge.net)
```
> sudo apt-get install libglew-dev
```

**1.3.6 -** Install the freeGLUT library (http://freeglut.sourceforge.net)
```
> sudo apt-get install freeglut3-dev
```

**1.3.7 -** Install the GIT version control system (https://git-scm.com)
```
> sudo apt-get install git
```

**1.3.8 -** Install the X11 tests programs (xeyes, xclock, xcalc, etc.)
```
> sudo apt-get install x11-apps
```

**1.3.9 –** Adjust the environment variable DISPLAY
```
> export DISPLAY=:0
```
Try to open a X11 window:
```
> xeyes
```
If this works, proceed to step 2. If the error is `failed to open the display', try a different DISPLAY value.
Find the IP address of the WSL2 machine:
```
> cat /etc/resolv.conf
```
This should return the IP number, e.g. 'nameserver 10.16.0.7'.
Use this IP number instead of X.X.X.X below:
```
> export DISPLAY= X.X.X.X:0
> xeyes
```

If this works, proceed to step 2.
If 'play' fails to 'open the display', try to run it from within a 'xterm' window:
> `sudo apt install xterm`
> `xterm –display :0`
In the new window that opens:
> `xeyes`

## 2 - Installation of Cytosim

These steps are similar for any OS: MacOS, WSL and Linux.
In the following, commands will be issued in the terminal, and the 'current working directory' should be set to the directory in which the simulation was compiled. This directory will be referred to as being the *base directory*. Alternatively, everything can be done in a separate directory, if files are copied as needed. If unfamiliar with the command line, consider following a tutorial, for example:
   https://www.learnenough.com/command-line-tutorial
   https://learnpythonthehardway.org/book/appendixa.html

### 2.1 - Download Cytosim source code
> `git clone https://gitlab.com/f-nedelec/cytosim.git cytosim`

Note: a Gitlab account is not necessary to download the code.
This should create a new subdirectory 'cytosim' in the current directory.

### 2.2.1 - Compile
> `cd cytosim`
> `make`

This should create 3 files in a subdirectory 'bin'. Check by running:
> `ls bin`

### 2.2.2 - Alternative compilation route
If step 2.2.1 failed, try this alternative method:
   Install cmake (https://cmake.org)
> `mkdir b`
> `cd b`
> `cmake ..`
> `make`
> `cd ..`

### 2.3 - Check executables
> `bin/sim info`
> `bin/play info`

Verify that both executables have been compiled to perform 2D simulations.
The output of the 'info' queries above should include one line saying 'Dimension: 2':
   **Dimension: 2** Periodic**:** 1 Precision**:** 8 bytes
*If this is not the case, and only then*, modify the file 'src/math/dim.h' (open, edit to change DIM to 2, save) and recompile Cytosim:
> `make clean`
> `make`

### 2.4 - Test run
**2.4.1 -** Copy three executables

```
> cp bin/sim sim
> cp bin/play play
> cp bin/report report
```

**2.4.2 -** Create a new directory
```
> mkdir run
```

**2.4.3 -** Copy and rename standard configuration file
```
> cp cym/fiber.cym run/config.cym
```

**2.4.4 -** Start simulation
```
> cd run
> ../sim
```

**2.4.5 -** Visualize the results of the simulation
```
> ../play
```
Press the space bar to animate. Press 'h' for help on keyboard shortcuts.
To quit the program, on MacOS try ⌘-Q , or CTRL-Q, or select Quit from the menu.
As a last resort, you can enter CTRL-C in the terminal window from which you started 'play'.

**2.4.6 -** Run in live mode
```
> ../play live
```

## 3. Configuration of the simulation

### 3.1 - Install editor
Install a code-oriented text editor (TextMate, SublimeText, etc.) if not already available.

### 3.2 - Open the editor
Start the code editor and open the file 'config.cym', located in the 'run' directory created in **2.4.2**.
This is a copy of 'fiber.cym'. Familiarize yourself with the different sections of the configuration file.
A parameter table is provided in the supplementary material.

### 3.3 - Modify the system to be simulated

**3.3.1 -** In the editor, make the following changes to 'config.cym':
```
radius=5 → radius=3
new 1 filament → new 100 filament
```
The radius of the circle is here specified in μm, like all distances in Cytosim. Cytosim follows a system of units adapted to the scale of the cell, derived from: micrometers, piconewton and seconds.

**3.3.2 -** Save the file without changing its name or location (overwrite 'config.cym')

**3.3.3 -** Check that the simulation is modified as expected
Switch to the Terminal Window, and enter:
```
> ../play live
```

### 3.4 - Modify the system further by editing 'config.cym'

**3.4.1 -** Adjust the filament's properties
In the paragraph 'set fiber filament', make the following changes:
    rigidity=20 → rigidity=0.1
    segmentation=0.5 → segmentation=0.2
    confine=inside, 200, cell → confine=inside, 10, cell
The bending rigidity is specified in pN µm$^2$. The segmentation is the approximate distance between the vertices describing the filaments, in µm. It needs to be reduced as filaments are made more flexible. The last change reduces the stiffness (pN/µm) of the confinement potential associated with the 'cell' edges.

**3.4.2 -** Shorten the filaments and adjust their initial position
In the paragraph 'new filament', make the following change:
    length=12 → length=2
All filaments will now be created with a length of 2 µm.

**3.4.3 -** Sign the configuration file, by editing the first line:
    % F. Nedelec, April 2010 → % Your name, date, place

**3.4.4 -** Check the validity of the configuration file (repeat **3.3.2** and **3.3.3**)

### 3.5 - Create passive connectors

**3.5.1 -** Define a molecular activity with affinity to the filaments
Add a new paragraph to 'config.cym', before the lines with the 'new' and 'run' command:
```
set hand binder
{
    binding_rate = 10
    binding_range = 0.01
    unbinding_rate = 0.2
}
```
In Cytosim, rates are specified in units of s$^{-1}$, and the range in µm. The order of the commands in the configuration file is important, and this paragraph must appear before any 'run' command, otherwise it will not be effective.

**3.5.2 -** Create bi-functional entities with these 'binders'
Add a paragraph to 'config.cym', following the previous one:
```
set couple crosslinker
{
    hand1 = binder
    hand2 = binder
    stiffness = 100
    diffusion = 10
}
```
The stiffness is specified in units of pN/µm, and diffusion coefficient in µm$^2$/s.
To learn more about Cytosim's objects, follow the tutorial in 'tuto_introduction.md', which is in a subfolder 'doc' of the base directory: doc/tutorials/

**3.5.3 -** Add the crosslinkers to the system
Add a new line after the paragraphs added in **3.5.1** and **3.5.2**, and before the 'run' command:
    new 1000 crosslinker

**3.5.4 -** Check the validity of the configuration file (repeat **3.3.2** and **3.3.3**)

### 3.6 - Create bifunctional motors

**3.6.1 -** Define a molecular motor activity
Add a new paragraph to 'config.cym', before the lines with the 'new' and 'run' command:
```
set hand motor
{
    binding_rate = 10
    binding_range = 0.01
    unbinding_rate = 0.2
    activity = move
    unloaded_speed = 1
    stall_force = 3
}
```
The 'activity = move' specifies a motor that moves continuously over the filament. The motor obeys a linear force-velocity curve (Figure 1), characterized by the unloaded speed (µm/s) and the stall force (pico Newton). At each time-step $\tau$, it moves by a distance $v \times \tau$. Using 'activity = none' would specify a molecule that does not move, and since this is the default activity, one could simply omit 'activity' altogether. Other activities are possible, for example 'activity = walk' would specify a motor with discrete steps along the filament.
The 'unloaded_speed' is positive for a plus-end directed motor. Specifying a negative value would make the motor move towards the minus-ends. This is further explained in Cytosim's first tutorial (see link in **3.5.2**).

**3.6.2 -** Create bi-functional entities with two `motors`
Add another paragraph to 'config.cym', following the previous one:
```
set couple complex
{
    hand1 = motor
    hand2 = motor
    stiffness = 100
    diffusion = 10
}
```

**3.6.3 -** Add the motors to the system
Add a new line after 'new filament', and before the line with the 'run' command:
```
new 200 complex
```

**3.6.4 -** Check the validity of the configuration file (repeat **2.3.2** and **2.3.3**)

### 3.7 - Reduce the time to be simulated

**3.7.1 -** In the editor, make the following changes to 'config.cym':
```
run 5000 system  →  run 1000 system
```
**3.7.2 -** Save the file

**3.7.3 -** Run the simulation
```
> ../sim
```
Note how much time was needed to compute this simulation. In the following sections, this time will be multiplied by the number of simulations made to generate a graph where a parameter is varied.

**3.8 -** Visualize the results
```
> ../play
```

**3.9 -** Copy the final configuration file to the base directory
```
> cp config.cym ../config.cym
```

## 4. Parameter Sweep
In this section, the number of crosslinkers in the network is systematically varied.

**4.1 - Import Python scripts:**
Create subdirectory 'byn', inside the base directory
```
> mkdir byn
```
Copy three scripts from the standard Cytosim distribution:
```
> cp python/run/preconfig.py byn/preconfig
> cp python/look/scan.py byn/.
> cp python/look/make_page.py byn/.
```

**4.2 - Make the files executable**
```
> chmod +x byn/preconfig
> chmod +x byn/scan.py
> chmod +x byn/make_page.py
```
Check that the scripts execute correctly:
```
> byn/preconfig help
> byn/scan.py help
> byn/make_page.py help
```
This should print a description of how each command can be used. It is possible to modify the PATH variable to call the scripts more simply, see https://en.wikipedia.org/wiki/PATH_(variable).

**Troubleshooting**:
If the error is 'command not found', check the path specified, which should correspond to the location of the file. Another type of error arises, if the OS does not provide python2. In that case, edit the 3 .py files and modify the shebang, which is the first line of each file (just add '3'):
```
#!/usr/bin/env python → #!/usr/bin/env python3
```

**4.3 - Copy configuration file to make it a 'template'**
```
> cp config.cym config.cym.tpl
```

**4.4 - Edit template configuration file**
In the code editor, open the file 'config.cym.tpl', located in the base directory.
Change one line, to introduce a variable text element:
```
new 1000 crosslinker → new [[range(0,4000,100)]] crosslinker
```

Preconfig will recognize the code contained in the double brackets, and replace it by a value obtained by executing this code in Python. In Python `range(X,Y,S)` specifies integers from X to Y, with an increment of S; in this case [0, 100, 200, 300, ... 4000]. The intent is to change the number of crosslinkers added to the simulation.

**4.5 - Generate config files from the template**

```
> byn/preconfig run%04i/config.cym config.cym.tpl
```
This should create 40 files, one for each value specified in `range(0,4000,100)'

Note: the names of the files to be generated by Preconfig are specified by 'run%04i/config.cym'. With this specific code, Preconfig will generate 'run0000/config.cym', 'run0001/config.cym', etc. It will actually make the directories 'run0000', 'run0001' etc., and create a file 'config.cym' in each of them. Please check Preconfig's help for more details (you can get this help by running `byn/preconfig help`).

**Troubleshooting**:
In case of error, check that the command is type exactly. Every character counts!

**4.6 - Run all simulations sequentially**
```
> byn/scan.py '../sim' run????
```

The python program 'scan.py' will execute the quoted command in the list of directories provided. In this case, this list of directories is specified by 'run????'. The question mark is a wild-card that matches any one character. Hence 'run????' matches any name starting with 'run' and followed by exactly four characters. Alternatively, one could use 'run*' to generate the same list, since '*' matches any string.

Note: in case of problems it can help to execute this command from a new terminal window.

The command will run 'sim' sequentially into all the directories created in step **4.5**. This may take half an hour to complete, but this depends very much on the computer capacities.
The amount of calculation requested can be reduced by editing 'config.cym':
```
run 1000 system → run 500 system
```

**4.7 - Visualize some simulations**
```
> ./play run0010
> ./play run0020
> ./play run0030
```

**4.8 - Generate images for all simulations**
```
> byn/scan.py '../play image size=256 frame=10' run????
```

**4.9 - Generate an HTML summary page**
```
> byn/make_page.py tile=5 run????
```
Open 'page.html' in a web browser

# 5. Making a graph
In this section, a plot is made from the results of the parameter sweep.

**5.1 -** Check the *report* executable:
```
> cd run
> ../report network:size
```
For each frame in the trajectory, this will print some comments, and two numbers: 'polymer surface'
Only the network 'surface' (last column) will be used.
It is possible to specifically restrict the information to one frame:
```
> ../report network:size frame=10
```
Comments can be removed, to make the output easier to process:

```
            > ../report network:size frame=10 verbose=0
```
Finally, the output can be redirected to a file using the Unix pipe facility ('>'):
```
            > ../report network:size frame=10 verbose=0 > net.txt
```
Print the content of the file to terminal for verification:
```
            > cat net.txt
```
This should print two numbers. The second one is the surface covered by the network in the 10$^{th}$ frame.
```
            > cd ..
```

**5.2 -** Generate a report in each simulation subdirectory:
```
            > byn/scan.py '../report network:size frame=10 verbose=0 > net.txt' run????
```

Again, scan.py will execute the quoted command in all directories specified by 'run????'. The command is the last one tried in **5.1**. This will generate a file 'net.txt' in every directory (try `'ls run????'`)

**5.3 -** Examine these numbers:
```
            > byn/scan.py 'cat net.txt' run????
```
Add the '+' option to concatenate the name of the directory and the command's output:
```
            > byn/scan.py + 'cat net.txt' run????
```

**5.4 -** Collect these numbers into a file
```
            > byn/scan.py + 'cat net.txt' run???? > results.txt
```

**5.5 -** Clean the file 'results.txt' by deleting the repeated text that is not numeric
  Open 'result.txt' in the code editor, and use the 'replace text' to remove all 'run' strings.
  Remove only 3 letters, and keep the numbers: 'run0000' should become '0000'.
  This should leave a file with only numbers organized in three columns.

**5.6 -** Generate a *contractility plot* from the data in 'result.txt'
  Use any method to generate a plot using column 1 for X and column 3 for Y.
  Label the X-axis as: number of crosslinkers / 100
  Label the Y-axis as: surface (micrometer^2)

**5.7 -** Store all simulations in a separate directory
```
        > mkdir save1
        > mv run???? save1
```

## 6. Alternative method for making a graph

**6.1 -** Add a report command at the end of 'config.cym.tpl'.
It should be on a new line after the closing '}' of the run command:
```
        report network:size net.txt { verbose = 0 }
```

**6.2 -** Generate config files
```
            > byn/preconfig run%04i/config.cym config.cym.tpl
```

**6.3 -** Run all simulations, using parallel threads:
```
            > byn/scan.py '../sim' run???? njobs=4
```
The number of jobs (njobs=4) should be adjusted, depending on the computer's capacity.
Monitor the usage of the CPU resources, for example by Activity Monitor App on MacOS.

**6.4 -** Collect the data
```
> byn/scan.py + 'cat net.txt' run???? > results.txt
```

**6.5 -** Make a plot.: same as **5.5** and **5.6**

**6.6 -** Store all the previous run in a separate directory
```
> mkdir save2
> mv run???? save2
```

## 7. Improved plot using random sampling

A variable is sampled here using a generator function from the 'random' module of Python.

**7.1 -** Edit the template configuration file
In the code editor, open the file 'config.cym.tpl', located in the base directory.
Add one line, just before the 'new crosslinker', to create a random variable:
```
[[num = int(random.uniform(0,4000))]]
```
Change one line, to introduce a variable text element:
```
new [[range(0,4000,100)]] crosslinker → new [[num]] crosslinker
```
Below this line, add another line to print the parameter value:
```
%[[num]] xlinkers
```

**Note 1:** the '%' character is important, since it will instruct Cytosim to skip the line.
Soon 'xlinkers' will serve as a recognizable tag, and it must appear only once in each file.

**Note 2**: At the end of 'config.cym.tpl', after the 'run {...}' command, there should be a report instruction:
```
report network:size net.txt { verbose = 0 }
```

**7.2 -** Generate config files
```
> byn/preconfig run%04i/config.cym 42 config.cym.tpl
```
This should generate 42 files. This number can be increased if the computer is fast enough.

**7.3 -** Check the values generated randomly:
```
> grep xlinkers run????/config.cym
```

**7.4 -** Run all simulations in parallel
```
> byn/scan.py '../sim' run???? njobs=4
```

Note: adjust the number of jobs to the computer's capacity. For best results, njobs should be set equal to the number of cores of the computer.

**7.5 -** Collect the data
To collect data from two different files, in each run directory, a solid strategy would be to write a Python script to read the files and reformat the data. However, in this case, a single command will suffice:
```
> byn/scan.py + 'grep xlinkers config.cym; cat net.txt' run???? > results.txt
```

The '+' option of scan.py will concatenate all the output of the command to a single line.
In this case, the command is actually composed of two commands separated by a semi-column (';').

**7.6 -** Clean the file
Open 'result.txt' in the code editor. All the data corresponding to one simulation should be contained on one line. Delete all occurrences of the string 'xlinkers'. Also remove the '%' characters at the start of each line. Finally save 'result.txt' which should be neatly organized in three columns of numeric values.

**7.7 -** Make a graph to plot the data:
    Same as **5.6**, but label the X-axis as: number of crosslinkers

**7.8 -** Save the dataset into a new subdirectory
```
> mkdir set1
> mv run???? set1/.
> mv results.txt set1/.
```

**7.9 -** Calculate another dataset:
    Repeat steps **7.2** to **7.6**.
    Repeat **7.8**, replacing `set1` by `set2` in all three commands

**7.10 -** Make combined plot:
    Concatenate the two data files:
```
> cat set?/results.txt > results.txt
```
    The new 'result.txt' should be larger. Plot data to make a combined graph, as described in **7.7**

## 8. Fully automated pipeline

In this part, all operations requiring manual intervention are replaced by commands. When this is done, it will be possible to write a single script to performs all the steps automatically. This script can be run on a remote computer such as a compute farm.

**8.1 –** Use the Unix 'sed' command to 'clean' the results.txt file, replacing step 7.6:

**8.1.1 –** Remove the '%' at the beginning of lines:
```
> sed –e 's/%//g' results.txt > tmp.txt
```

**8.1.2 –** Remove 'xlinkers':
```
> sed –e 's/xlinkers//g' tmp.txt > results.txt
```

**8.2 –** Write a script to call all the commands successively in the right order
    There are different ways to do this, using different languages: bash or Python.
    create a file 'pipeline.bash', and copy all the commands needed to run the pipeline
    Execute the script:
```
> bash pipeline.bash
```

**8.3 –** Use an automated plotting script
    It can be advantageous to remove all human
    Many tools exist to generate a PDF plot directly from a data file, for example:
    gnuplot (http://www.gnuplot.info), PyX (https://pyx-project.org) or Seplot (https://pypi.org/project/seplot/)

**REPRESENTATIVE RESULTS:**

**In section 2:**
Successful compilation of Cytosim using 'make' should produce sim, play and report in the subdirectory 'bin'. The output of step 2.3 (`sim info`) should be similar to:

```
─────────────────────────────────────────────────────────
|  CytoSIM 2D  —  www.cytosim.org  —  version PI  — June 2019  |
─────────────────────────────────────────────────────────
   Dimension: 2    Periodic: 1    Precision: 8 bytes
   Fiber lattice 0
   Built Mar 22 2022 09:16:52 with 9.4.0
   Code version a838c78 with assertions
```

**In section 3:**
The configuration file should be similar to jove.cym, provided as supplementary material.

**In section 4:**
Images obtained in step 4.8 should be similar to the one shown on Figure 2.
The HTML summary page obtained in step 4.9 is shown on Figure 3.

**In sections 5 & 6:**
Expected plots are on Figure 4.

**In section 7:**
Expected plots are on Figure 5.

**FIGURE AND TABLE LEGENDS:**

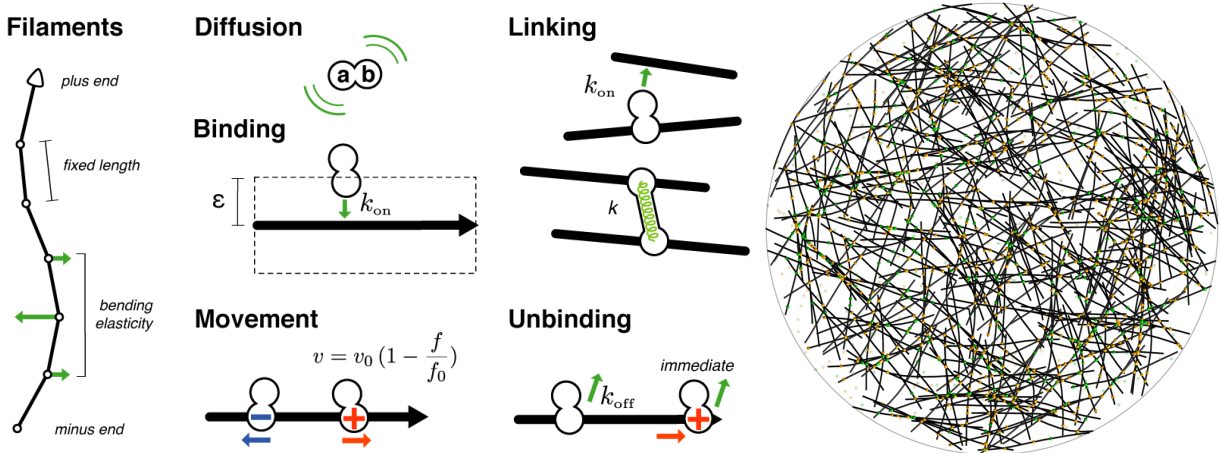

**Figure 1: Left:** some of Cytosim's modelling assumptions. Filaments are represented by line segments of identical length joined to account for bending elasticity. Connectors are composed of two units (a, b) linked together by a Hookean spring. Units can bind to filaments that are within their binding range $\varepsilon$, with a prescribed binding rate $k_{on}$. They unbind spontaneously at a rate $k_{off}$ or upon reaching the end of the filaments. Connectors made of two units can connect two filaments. These units can represent different activities. In this work 'binders' are passive elements that bind/unbind without moving and 'motors' actively move along filaments. They move towards the plus ends of the filament, and the force opposed to this movement affects the speed of translocation, following the equation shown. The speed $v$ of the motors obeys a linear force-velocity curve, characterized by the unloaded speed $v_0$ (µm/s) and the stall force $f_0$ (pN). Any opposing force $f$ would reduce the speed of the movement. **Right:** Network simulation developed in this protocol. Black lines represent the filaments. Green and orange dots represent the motor and the crosslinkers.

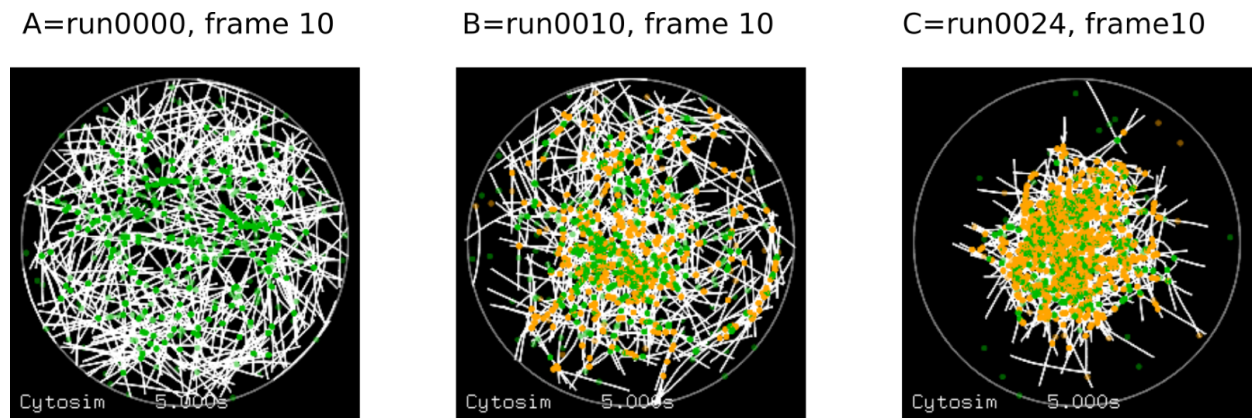

**Figure 2: Graphical representation of the networks.** Three different systems represented here at the same time (t=5s). The three simulations (left, center, right) contain the same amount of motors (green) but different amounts of crosslinkers (orange) and contract at different rates, as can be seen from the area reached after 5s.

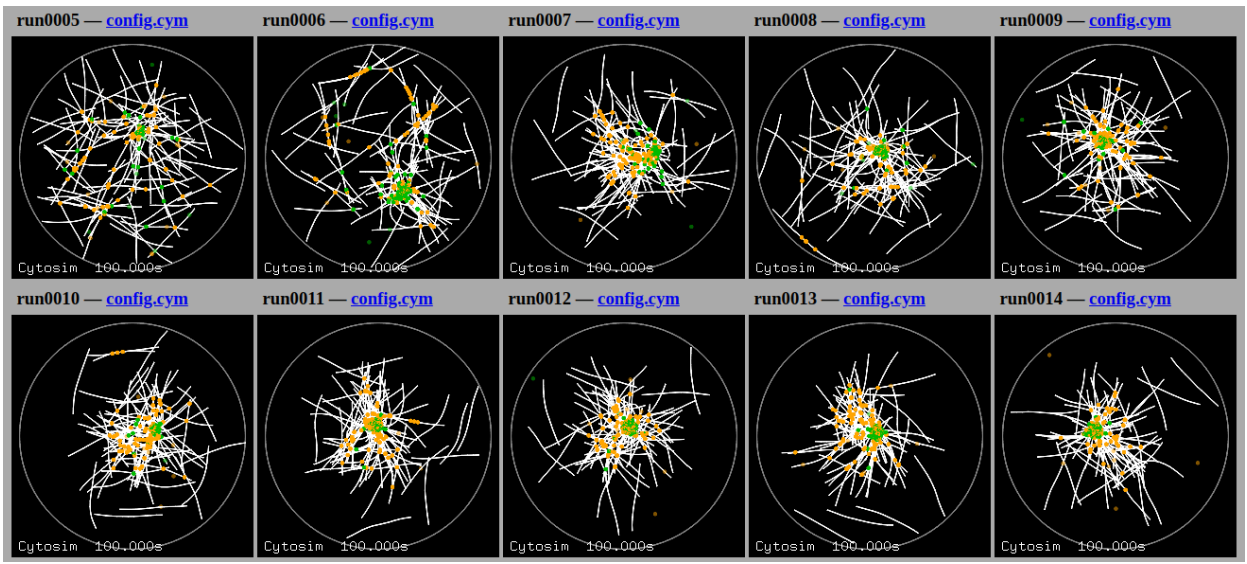

**Figure 3: Graphical representation of ten networks.** Ten different simulations represented at the same time point (t=100s). The systems differ by their composition, specifically the number of crosslinkers. The exact value can be obtained by clicking on the 'config.cym' link located above each simulation, which should open the file containing all parameters. The file is contained in the directory as indicated in black above each image.

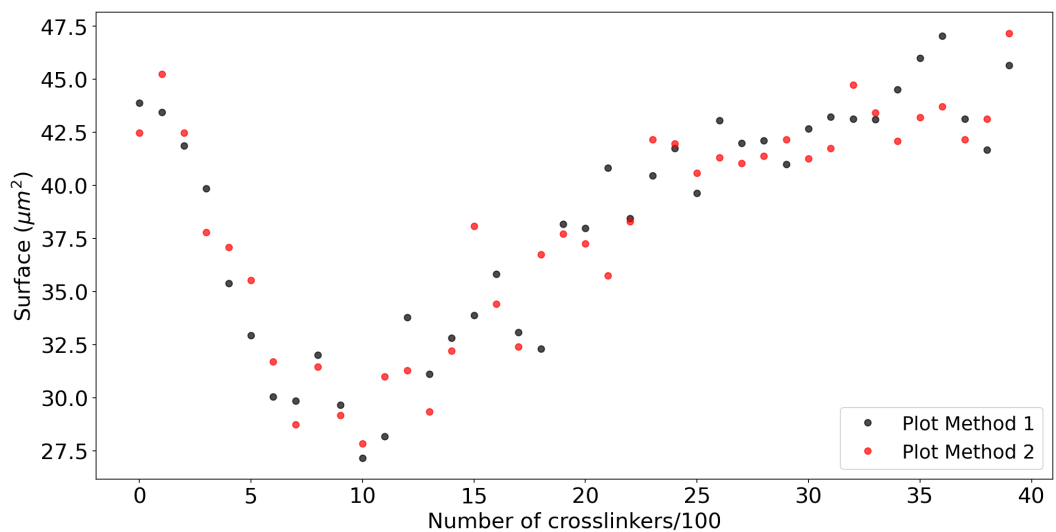

**Figure 4: Contraction plot from regular parameter sampling.** This plot is obtained in step 5 and 6 of the protocol. The number of crosslinkers in the system is varied regularly (X-axis). The resulting contractility (Y-axis) is obtained directly from the coordinates of the vertices describing the filaments (see **5.1**).

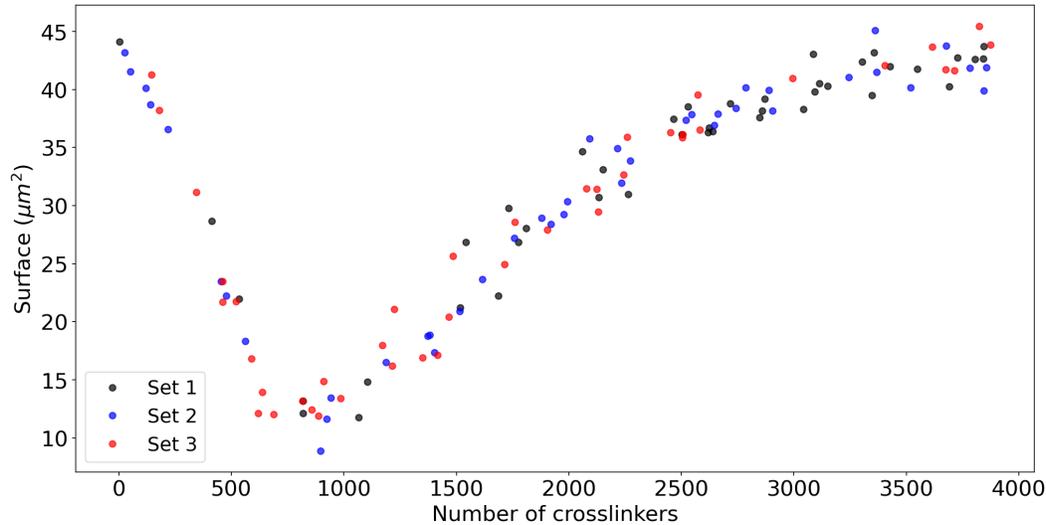

**Figure 5: Contraction plot from randomized sampling of the parameter.** This kind of plot is obtained in step 7 of the protocol. The plot is as described on Figure 4, but points are randomly distributed along the X-axis.

**TABLE OF MATERIALS:**

| Cytosim source code | https://gitlab.com/f-nedelec/cytosim.git |
|---|---|
| preconfig | https://github.com/nedelec/preconfig |
| scan.py | https://github.com/nedelec/scan.py |
| make_page.py | https://github.com/nedelec/make_page.py |
| jove.cym | included as supplementary material |

**DISCUSSION:**

The method outlined in this article relies on three small and independent python programs, which were used in diverse ways along the described protocol. The first script *preconfig* is a versatile tool that can replace the need of writing custom python scripts[27]. It is used to generate multiple configuration files from a single template file, specifying which parameter should be varied, and how it should be varied. To vary multiple parameters, simply add more code snippets into the template file, and *preconfig* will generate all the possible combinations. One advantage of the template method is that all the information about the variations (which parameters are varied and how) is contained in a single file, with a minimal amount of code. The script 'scan.py' is also versatile and is used to execute commands into subdirectories. Finally, 'make_page.py' produces a HTML page to view the results of multiple simulations side-by-side. This takes advantage of the fact that browsers have been highly optimized to display panels of images. This script becomes particularly useful when multiple graphs are made (e.g. using scan.py) in the simulations subdirectories. There are many alternatives to the approach described here, and some programmers script in bash or prefer to write C programs to perform these tasks. Python however offers the advantage of a clean and modern syntax, and, being interpreted, Python scripts often work 'out of the box' on modern architecture, including most high-performance clusters. The most significant drawback is that since Python is constantly evolving, it is possible that these scripts malfunction in a different version of Python. It is important, here as well, to use the latest versions, and to report back any trouble.

Varying a parameter is a standard approach to determine the influence that this parameter may have on a system. This technique requires setting a range over which to perform the variations. This range should be chosen according to what is believed to be realistic *in vivo*. When this is not known, physical or numerical considerations can help to narrow down the choice. For example, if the total simulation time is T, then it is probably futile to probe rates (R) corresponding to exceedingly rare events, such that R×T << 1. In general, multiple parameters are varied, and the different ranges can be refined iteratively. Upon examination of the previous results, one may choose a region of parameter space where interesting transitions are occurring. It is essential to quantify such occurrences using a metric consisting of a few numerical values (ideally one) to characterize the behavior of the system. In the case of the actomyosin network, a natural choice was to estimate the surface of the network after a fixed amount of 'simulated time' (achieved at frame 10) to monitor the contractility of the system. Given that the network is homogeneous and covers a disc, the surface $S$ covered by the network was estimated from the filament's $P$ vertices positions $x_i$:

$$S = \frac{2\pi}{P}\sum_i (x_i - c)^2 \quad \text{with} \quad c = \frac{1}{P}\sum_i x_i \quad \text{the center of mass.}$$

This choice was justified previously[11,24]. Finding a good metric that is simple, robust and biologically relevant will impact the entire study. This calculation is done by Cytosim's report 'network:size' command, and is thus the quantity plotted on the Y-axis of all graphs. Implementing new report calculations requires only basic knowledge of C++ and Cytosim's code structure. Occasionally, it can be more convenient to use a Python script to calculate a metric, from quantities exported by Cytosim, in particular if a python module already does part of the job.

Sweeping a parameter using random variables (Section 6) offers several advantages compared to a regular variation (Section 3). Firstly, it does not require one to decide how many values will be sampled. In fact, at any time, more points can be added to the dataset. Similarly, missing data points will be unnoticed, which is not the case of regular variations. Large computing clusters are not always reliable, and the failure of a computer node may result in missing of some data points. With regular sampling, one has to identify and reschedule the missing jobs and wait for their completion. The random sampling strategy avoids this issue. More importantly, perhaps, random sampling avoids correlation to be introduced unwillingly, when multiple parameters are varied. For example, if a parameter X is varied with factors 1, 2, 4, 8, etc. and a second parameter Y is also varied in the same way, then their ratio is varied in a more limited way. In other words, a regular grid in high dimensions can have many levels of hidden symmetries: diagonals, etc. With random sampling, any combination of parameters is also a random variable, and thus no apriorism is hidden in the sampling strategy. Another advantage is that random sampling permits the data to be split easily into pools of equivalent size, and thus variations can be estimated with a bootstrapping method, *a posteriori*. The method also has some drawbacks. Random sampling requires more simulations to cover a certain area of parameter space with the same precision, compared to regular sampling, because its points are not regularly spaced. Some amount of computation is wasted when two simulations are performed with nearly identical parameter values. Another shortcoming of random sampling is that, even with a high sampling rate, some area of the parameter space will remain unexplored. It is also more difficult to plot variations over two parameters, as a surface embedded in 3D or a color-coded 2D area, but this can still be done. Plots showing variation against only one parameter however are easy to make and offer more insights than equivalent plots obtained with regular sampling, because they include noise in all directions (compare the plots obtained in steps 4 and 6).

The 2D actomyosin network can be calculated rapidly. It uses ubiquitous elements of the cytoskeleton: filaments, molecular motors and crosslinkers. Cytosim can simulate many other systems, and different components can be included simply by adding extra commands to the configuration file. It is often possible to directly copy-paste from the many examples provided with the source code, which illustrate different classes of objects present in cells: microtubules, centrosomes, nucleus, etc. There are, of course, limitations to the degree of complexity that can be handled, by the software/hardware combination and by the investigator. Overly complex systems require longer calculation times, and are hard to explore numerically. However, nature offers us plenty of relatively small cytoskeletal systems for which the current software hardware combination is sufficient. The field is still in its infancy and while the future looks bright and open, the challenges are enormous and teams should be working together to develop the best software solutions. There are few alternatives to using computers outside bench work, when it comes to analyzing complex systems like the ones found in living cells. There are, however, many ways to use computers with this aim, and Cytosim only represents one possible approach. Cytosim has been in constant development for the past 20 years and although it is already a large-scale project; many aspects of the cytoskeleton are not covered. Yet, it has been used by many groups independently of its authors, and the resulting publications[28-37] are a clear testimony that the existing capabilities are sufficient for many research projects. These capabilities were often expended independently of Cytosim's main authors, which testifies that the code is modular and sufficiently clearly organized to permit this.

## ACKNOWLEDGMENTS

We thank members the SLCU modelling club, especially Tamsin Spelman, Renske Vroomans, Cameron Gibson, Genevieve Hines, Euan Smithers and other beta testers of the protocol: Wei Xiang Chew, Daniel Cortes, Ronen Zaidel-Bar, Aman Soni, Chaitanya Athale, Kim Bellingham-Johnstun, Serge Dmitrieff, Gaëlle Letort and Ghislain de Labbey. We acknowledge support from the Gatsby Charitable Foundation (Grant PTAG-024) and the European Research Council (ERC Synergy Grant, project 951430).

## DISCLOSURES:

The authors declare having no conflict of interest.

## SUPPLEMENTARY FILES:

| fiber.cym | Configuration file, used to start step 3 |
| --- | --- |
| jove.cym | Configuration file (obtained at the end of step 3) |
| config.cym.tpl | Template configuration file (step 4) |
| preconfig | Versatile file generator |
| scan.py | Python script to execute command in subdirectories |
| make_page.py | Python script to generate HTML summary pages |
| Parameter table | Description of the parameters values used in the simulation |